\documentclass{webofc}

\usepackage[varg]{txfonts}   
\usepackage{hyperref}
\usepackage{url}
\hypersetup{colorlinks=true,citecolor=blue,urlcolor=blue,linkcolor=blue}
\usepackage{bm}
\def\nslash{\rlap{\hspace{0.02cm}/}{n}}
\def\Dslash{\rlap{\hspace{0.07cm}/}{D}}
\begin{document}
\title{Recent developments in HQET}
%
%

\author{\firstname{Gil} \lastname{Paz}\inst{1}\fnsep\thanks{\email{gilpaz@wayne.edu}}        
}

\institute{Department of Physics and Astronomy,
Wayne State University, Detroit, Michigan 48201, USA}

\abstract{In this talk we review recent perturbative and non-perturbative developments in Heavy Quark Effective Theory (HQET).
}
\maketitle
\section{Introduction} \label{sec:intro}
Heavy Quark Effective Theory (HQET) is an effective field theory for heavy quarks. It is most useful for cases where the quark mass $m_Q$ is much larger than the QCD scale, namely, $m_Q\gg\Lambda_{\mbox{\scriptsize QCD}}$. Using well-known transformations, see, e.g., \cite{Neubert:1993mb, Manohar:2000dt}, we can obtain from the QCD Lagrangian  the following  Lagrangian: 
\begin{equation}\label{eq:QCD_to_HQET}
{\cal L}=\bar h_viv\cdot Dh_v+\bar h_vi\Dslash_\perp\frac1{2m_Q+iv\cdot D}i\Dslash_\perp h_v\,,
\end{equation}
where $h_v$ is the heavy quark field and $D^\mu_\perp=D^\mu-(v\cdot D)v^\mu$. For $v=(1,\vec{0})$, $D^\mu_\perp=\vec{D}$. Expanding in powers of $iv\cdot D/2m_Q$ gives the dimension-five HQET Lagrangian: 
\begin{equation}
{\cal L}_{\mbox{\scriptsize HQET}}=\bar h_viv\cdot Dh_v-c_2\bar h_v\frac{D^2_\perp}{2m_Q}h_v-c_F\bar h_v\frac{\sigma_{\alpha\beta}G^{\alpha\beta}}{4m_Q}h_v+{\cal O}\left(\frac1{m_Q^2}\right).
\end{equation}
Expanding to higher orders in $1/{m_Q}$ would give the HQET Lagrangian with tree-level Wilson coefficients. Starting at dimension seven there are other operators with ${\cal O}(\alpha_s)$ Wilson coefficients, see \cite{Manohar:1997qy}, where the Lagrangian up to dimension seven was presented.  The dimension-eight Lagrangian was presented in  \cite{Gunawardana:2017zix} and \cite{Kobach:2017xkw}. Ref. \cite{Gunawardana:2017zix} also presented a method to easily construct local HQET operators of, in principle,  any dimension.  

Using HQET, observables can be written schematically as a series
\begin{equation}
\mbox{Observable }=\sum^\infty_{n=0}\sum_j c_n^j(\mu)\frac{\langle\, O_n^j(\mu)\,\rangle}{m_Q^n}\,, 
\end{equation}
where $\langle\, O_n^j(\mu)\,\rangle\sim \Lambda^n_{\mbox{\scriptsize QCD}}$ and $\mu\sim m_Q$. The Wilson coefficients $c_n^j(\mu)$ are perturbative and the matrix elements $\langle\, O_n^j(\mu)\,\rangle$ are non-perturbative. Since $\alpha_s(\mu)$ becomes smaller for large $\mu$ and $\Lambda{\mbox{\scriptsize QCD}}/m_Q$ is small, we expect to achieve good precision with just a few terms.  To improve the precision we can 
calculate $c_n^j(\mu)$ to higher powers in $\alpha_s$ and/or include $\langle\, O_n^j(\mu)\,\rangle$ with larger $n$, assuming we can extract them from data or calculate them using  Lattice QCD. What $\langle\, O_n^j(\mu)\,\rangle$ do we usually encounter?

Strong interaction operators are made of quarks and gluons. These can be local, e.g.,  $\bar{q}(0)\cdots q(0)$, or non-local, e.g., $\bar{q}(0)\cdots q(tn)$, where $n$ is a light-cone vector ($n^2=0$). The general matrix element is of the form, $\langle f(p_f)|O_n^j(\mu)|i(p_i) \rangle$, where  $O_n^j(\mu)$ can be local or non-local and $p_i, p_f$  can be independent or not. Let us look at the most common options, ordered by increased complexity. 

A local operator between the vacuum and a state gives rise to a number: decay constant, e.g., $\langle0|\bar q\gamma^\mu\gamma_5h_v| P(v)\rangle=-i\sqrt{m_P}{f_P}v^\mu$. Decay constants will appear in section \ref{sec:perturbative}. A diagonal matrix element of a local operator gives rise to a number: HQET parameter, e.g., $\langle {\bar B}|\bar b\, \bm{\vec{D}^2}\,b|{\bar B}\rangle=2M_B{\mu_{\pi}^2}$. HQET parameters will appear in section \ref{sec:perturbative}. A non-diagonal matrix element of a local operator gives rise to a function: form factor, e.g., $\langle D(p_f)|\bar c\gamma^\mu b|\bar{B}(p_i)\rangle={ f_+(q^2)}(p_i+p_f)^\mu+{ f_-(q^2)}(p_i-p_f)^\mu
$, where $p_f-p_i=q$. Form factors will appear in section \ref{sec:form_factor}. A non-local operator between the vacuum and a state gives rise to a function, e.g., LCDA (Light-Cone Distribution Amplitude): $
\langle H_v|\bar h_v(0)\nslash\gamma_5\,[0,tn]\,q_s(tn)| 0\rangle=-iF(\mu)\int_0^\infty d\omega\, e^{i\omega t}{\phi_+(\omega,\mu)}$. LCDAs will appear in section \ref{sec:LCDA}. A diagonal matrix element of a non-local operator gives rise to a function, e.g., shape function:
${S(\omega)}=\int_{-\infty}^{\infty}\,dt\,e^{i\omega t}\langle \bar{B}(v)|\bar{b}(0)\,[0,tn]b(tn)|\bar{B}(v)\rangle/4\pi M_B$. Finally, non-diagonal matrix element of a non-local operator, e.g.  $\langle K^{(*)}(p_f)|\bar{s}_L(0)\gamma^\rho\,\cdots \,\tilde{G}_{\alpha\beta}b_L(tn)|B(p_i)\rangle$, gives rise to a non-local form factor, see \cite{Khodjamirian:2010vf}. We will not discuss shape functions or non-local form factors here. 

In the following we review recent perturbative and non-perturbative developments in HQET. Recent is defined to be the period of Spring 2022-Spring 2024. To allow for a broad overview, only some aspects of the papers are highlighted. More details can be found in the original papers. The rest of the talk is structured as follows. Section \ref{sec:perturbative} discusses recent perturbative developments. Section \ref {sec:LCDA} discusses recent developments related to LCDAs. Section \ref{sec:form_factor} discuses recent developments related to form factors. Section \ref{sec:other} discusses other recent developments. We conclude in section  \ref{sec:conclusions}.

\section{Recent developments in HQET: Perturbative}\label{sec:perturbative}
As mentioned in the introduction, we can improve theoretical predictions by calculating $c_n^j$ to higher orders in $\alpha_s$. Moreno calculated $c_3^{\mbox{\scriptsize Darwin}}$  to ${\cal O}(\alpha_s)$ for $\bar B\to X_u\, \ell\,\bar\nu_\ell$ total rate and leptonic invariant mass in \cite{Moreno:2024bgq}. In some cases the ``technology" improved to ${\cal O}(\alpha_s^4)$. For example, Takaura used the four-loop relation between the pole and $\overline{\mbox{MS}}$ masses to extracted HQET parameters from $B$ and $D$ meson masses \cite{Takaura:2022eel}. Lee and Pikelner calculated the four-loop HQET propagator in \cite{Lee:2022art}. This four-loop calculation was used by Grozin to find the four-loop HQET heavy to light anomalous dimension \cite{Grozin:2023dlk}. Looking at  \cite{Grozin:2023dlk} in more detail, the anomalous dimension can be used to calculate the ratio of the $B$ meson and $D$ meson decay constants $f_B/f_D$. Including the four-loop calculation Ref. \cite{Grozin:2023dlk} finds
\begin{eqnarray}
\hspace{-2em}\dfrac{f_B}{f_D}&=&\sqrt{\dfrac{m_D}{m_B}}\left(\dfrac{\alpha_s^{(4)}(m_c)}{\alpha_s^{(4)}(m_b)}\right)^{-\frac{\tilde{\gamma}_{j0}}{2\beta_0^{(4)}}}\!\!\!\bigg\{1+\cdot\cdot \alpha_s+\cdot\cdot \alpha_s^2+\cdot\cdot \alpha_s^3+[\sim 1 \mbox{GeV}]\left(\dfrac1{m_c}-\dfrac1{m_b}\right)+\cdot\cdot\!\!\bigg\}\nonumber\\
&=&0.669 \cdot (1 + 0.039 + 0.029 + 0.032 + [\sim 0.46])\,,
\end{eqnarray}
where the last term includes an estimate of power corrections. Ref. \cite{Grozin:2023dlk} comments on this equation that ``Convergence of the perturbative series is questionable [...]".
Numerically, without power corrections $f_B/f_D=0.736$, and with power corrections $f_B/f_D=1.04$. Comparing these to the lattice QCD value, $f_B/f_D = 0.896\pm 0.009$, Ref. \cite{Grozin:2023dlk} concludes that  ``The effect of the (poorly known) $1/m_{c,b}$ correction is large."

\section{Recent developments in HQET: Non-local matrix elements} \label{sec:LCDA}
Non local matrix elements arise in many processes, e.g., the  proton parton distribution function in hard QCD processes. In B decays such as $B\to K^*\gamma$ we encounter the 
$B$-meson LCDA. It is the Fourier transform of $\langle B|\bar{b}(0)\,
\dots[0,tn]q_s(tn)|0\rangle$. $B$-meson LCDA also arises when the B meson is in the \emph{final} state, see, e.g., in the decay $W,Z\to B+\gamma$ \cite{Grossman:2015cak}. Such processes were recently considered in \cite{Beneke:2023nmj} ($W^\pm\to B\pm\gamma$) and \cite{Ishaq:2024pvm} ($W^+\to B^+\ell^+\ell^-$). 

Beneke, Finauri, Vos, and Wei considered the process $W^\pm\to B\pm\gamma$ in \cite{Beneke:2023nmj} . It has  three scales: hard scale $Q\gg$ heavy quark scale $m_Q\gg$ QCD scale $\Lambda_{\mbox{\scriptsize QCD}}$. In \cite{Beneke:2023nmj} the QCD LCDA:
\begin{equation}
\langle H(p_H)| \bar{Q}(0) \nslash \gamma^5[0,tn]q(tn) |0\rangle = 
-i f_H n\cdot p_H \int_0^1 du\, e^{i u t n \cdot p_H} \phi(u;\mu) \,,
\end{equation}
was matched to a perturbative function convoluted with HQET LCDA:
\begin{equation}
    \langle H_v| \bar{h}_v(0) \nslash\gamma^5[0,tn]q_s(tn) |0\rangle = -i F_{\rm stat}(\mu) \,n\cdot v \int_0^\infty d\omega\, e^{i \omega t n\cdot v} \varphi_+(\omega;\mu) \,.
\end{equation} 
Such a factorization allows to resum  large logs between $\Lambda_{\mbox{\scriptsize QCD}}$ and  $m_Q$ and $m_Q$ and the hard scale $Q$. This paper also consider the evolution of the LCDA. Starting with HQET LCDA at soft scale $\mu_s = 1$ GeV, it is evolved in HQET to the matching scale $\mu$ and matched to $\phi(u)$. It is then  evolved in QCD to the hard scale $m_W$, see figure 6 of  \cite{Beneke:2023nmj}.  The branching ratio they find is \cite{Beneke:2023nmj}
\begin{equation}
\text{Br}(W\to B \gamma) = (2.58 \pm 0.21_{\rm in}\,^{+0.05}_{-0.08}\,_{\mu_h}\,^{+0.05}_{-0.08}\,_{\mu_b}\,^{+0.18}_{-0.13}\,_{\delta}\,^{+0.61}_{-0.34}\,_{\beta}\,^{+2.95}_{-0.98}\,_{\lambda_B}) \cdot 10^{-12} \,.
\end{equation}
The uncertainty from the low-scale HQET LCDA parameters $\lambda_B,\, \beta$ is large.

Ishaq, Zafar, Rehman, and Ahmed considered the process $W^+\to B^+\ell^+\ell^-$ in \cite{Ishaq:2024pvm}. Using the scale hierarchy $m_W\sim m_b\gg \Lambda_{\mbox{\scriptsize QCD}}$ they factorize the amplitude as 
\begin{equation}
{\mathcal M}=e~\bar \ell\gamma^\mu \ell \int_0^\infty \!\! d\omega\, T_{\mu}(\omega, m_b,q^2, \mu_F)
\Phi^+_B(\omega,\mu_F)+\mathcal{O}\left(m_b^{-1}\right),
\end{equation} 
where $T_\mu$ is the perturbative hard-scattering kernel. Calculating $T_\mu$ at ${\cal O}(\alpha_s)$ they find ``[...] the scale dependence in the NLO decay rates [...] gets largely reduced, particularly for relatively large  $\mu_F$." See figure 5 of  \cite{Ishaq:2024pvm}.  The theoretical prediction for the branching ratio is $\sim10^{-11}$ for electrons and muons. It is sensitive to the hadronic parameter $\lambda_B$ defined as
\begin{equation}
\dfrac1{\lambda_B}\equiv\int_0^\infty \dfrac{d\omega}{\omega}\phi_B^+(\omega)\,.
\end{equation}
See figure 6 of \cite{Ishaq:2024pvm}. Ref. \cite{Ishaq:2024pvm} uses $\lambda_B=0.35\pm0.15$\, GeV. Searching for $B^+\to\ell^+\nu_\ell\gamma$ Belle got $\lambda_B=0.36^{+0.25}_{-0.09}$\, GeV \cite{Belle:2018jqd}. Ref. \cite{Ishaq:2024pvm} concludes that observing $W^+\to B^+\ell^+\ell^-$  at the LHC could help constrain $\lambda_B$.

\section{Recent developments in HQET: Local non-diagonal matrix elements} \label{sec:form_factor}
In the SM, $B\to D$ transitions are described by two form factors,  and $B\to D^*$ transitions are described by four form factors. At leading power in the heavy quark symmetry \emph{all} of these form factors are described by one universal Isgur-Wise function $\xi$. Including $1/m_c\,\&\,1/m_b$ power corrections there are three additional functions and the number grows rapidly at higher powers. Including $1/m_c^2$ corrections, there are additional 20 functions and including $1/m^2_c\,\&\,1/m^2_b$ there are additional 32 functions, see Table 1 of \cite{Bernlochner:2022qow}.

Bernlochner, Ligeti, Papucci, Prim, Robinson, and Xiong, suggested in \cite{Bernlochner:2022ywh} supplemental power-counting  that reduces these numbers. The QCD Lagrangian before the $1/m_Q$ expansion is given in equation (\ref{eq:QCD_to_HQET}). The postulated power counting is in powers of  $i\Dslash_\perp$: currents involve one $i\Dslash_\perp$, Lagrangian insertions involves two $i\Dslash_\perp$'s. Many subleading contributions arise from Lagrangian insertions. Ref.  \cite{Bernlochner:2022ywh}  conjectures that terms entering at third order or higher should be suppressed and calls this residual chiral (RC) expansion. Under RC expansion including $1/m_c^2$ corrections, there is one additional function and including $1/m^2_c\,\&\,1/m^2_b$ there are additional 3 additional functions, see Table 1 of \cite{Bernlochner:2022qow}. Even more recently, Bernlochner, Papucci, and Robinson, applied the same method to $\Lambda_b \to \Lambda_c l \nu$ decay \cite{Bernlochner:2023jkp}. 
 
 Turning to phenomenology recall that 
 \begin{equation}
 R(D^{(*)})\equiv{\mbox{Br}(\bar B\to D^{(*)}\, \tau\,\bar\nu_\tau)}/{\mbox{Br}(\bar B\to D^{(*)}\, \ell\,\bar\nu_\ell)}, \quad \ell=e,\mu
 \end{equation}
 Ref.  \cite{Bernlochner:2022ywh} obtained  $R(D)=0.288(4)$. This should be compared to HFLAV 2024  Standard Model (SM) prediction of $R(D) = 0.298(4)$, and value from experiment  of $R(D) = 0.342(26)$ \cite{HFLAV:Moriond 2024}. Ref.  \cite{Bernlochner:2022ywh} obtained  $R(D^*)=0.249(3)$. This should be compared to HFLAV 2024  SM prediction of $R(D^*) = 0.254(5)$, and value from experiment of $R(D^*) = 0.287(12)$ \cite{HFLAV:Moriond 2024}. Clearly the tension between experiment and theory remains for both theoretical predictions.

\section{Recent developments in HQET: Other topics}\label{sec:other}
Manzari and Robinson suggested a new theoretical framework for heavy quark resonances in \cite{Manzari:2024nxr}.  The new framework uses on-shell recursion techniques to express resonant amplitude as a product of on-shell  sub-amplitudes. In figure 5 of \cite{Manzari:2024nxr} the authors present a toy example calculation in this framework and a fixed-width Breit-Wigner side by side with Belle data for  $D_2^*$ resonance. From the figure the preliminary results of Ref. \cite{Manzari:2024nxr} seem promising.

Garg and Upadhyay studied F-wave B mesons in HQET in \cite{Garg:2022bhj}.  F-wave B mesons have angular momentum $L=3$. Adding  $L=3$ to the heavy-quark spin $s_Q=1/2$ gives 7/2 and 5/2 angular momenta. Adding the light quark spin 1/2 to 5/2 gives a $J=2$ and $J=3$ doublet. Adding the light quark spin 1/2 to 7/2 gives a $J=3$ and $J=4$ doublet. Thus the spectrum contains two doublets. Ref. \cite{Garg:2022bhj} used information from, e.g., D mesons, to calculate B meson properties. For example, in table 2 of \cite{Garg:2022bhj} the authors compare the calculated masses to two quark models. 

Vishwakarma and Upadhyay presented an analysis of 2S singly-heavy baryons in HQET in \cite{Vishwakarma:2022vzy}. They used information from measured 2S baryons: $\Xi_c(2970)$ and $\Lambda_b(6070)$, and HQET to calculate 2S baryon properties.  For example, in table 7 of \cite{Vishwakarma:2022vzy} the authors compare the calculated masses to two quark models.

\section{Conclusions}\label{sec:conclusions}
Arguably, this year marks the 35th anniversary of HQET, based on the date of the publication of ``Weak Decays of Heavy Mesons in the Static Quark Approximation,'' by Isgur and Wise \cite{Isgur:1989vq}. This paper includes the line ``This logarithm can be displayed explicitly by going over to an effective theory where the heavy quark is treated as a static color source." Note the words ``effective theory".

HQET is a mature field where some perturbative corrections are known to fourth order, and  some non-perturbative power corrections are known to fourth order. In this talk we reviewed   recent developments in HQET both on the perturbative side and the non-perturbative side. Theoretical progress in the field mirrors the experimental progress. It is clear from several of the papers discussed in this talk that non-perturbative effects often dominate the uncertainties. Controlling them might require further developments in HQET.  

\section*{Acknowledgements}
I thank the organizers for the invitation to give this talk. This work was supported by the U.S. Department of Energy grant DE-SC0007983.


\begin{thebibliography}{99}
%
%
\bibitem{Neubert:1993mb}
M.~Neubert,
Phys. Rept. \textbf{245}, 259-396 (1994)
[arXiv:hep-ph/9306320 [hep-ph]].

\bibitem{Manohar:2000dt}
A.~V.~Manohar and M.~B.~Wise,
Camb. Monogr. Part. Phys. Nucl. Phys. Cosmol. \textbf{10}, 1-191 (2000)

\bibitem{Manohar:1997qy}
A.~V.~Manohar,
Phys. Rev. D \textbf{56}, 230-237 (1997)
[arXiv:hep-ph/9701294 [hep-ph]].

\bibitem{Gunawardana:2017zix}
A.~Gunawardana and G.~Paz,
JHEP \textbf{07}, 137 (2017)
[arXiv:1702.08904 [hep-ph]].

\bibitem{Kobach:2017xkw}
A.~Kobach and S.~Pal,
Phys. Lett. B \textbf{772}, 225-231 (2017)
[arXiv:1704.00008 [hep-ph]].

\bibitem{Khodjamirian:2010vf}
A.~Khodjamirian, T.~Mannel, A.~A.~Pivovarov and Y.~M.~Wang,
JHEP \textbf{09}, 089 (2010)
[arXiv:1006.4945 [hep-ph]].

\bibitem{Moreno:2024bgq}
D.~Moreno,
Phys. Rev. D \textbf{109}, no.7, 074030 (2024)
[arXiv:2402.13805 [hep-ph]].

\bibitem{Takaura:2022eel}
H.~Takaura,
EPJ Web Conf. \textbf{274}, 03003 (2022)
[arXiv:2212.02874 [hep-ph]].

\bibitem{Lee:2022art}
R.~N.~Lee and A.~F.~Pikelner,
JHEP \textbf{02}, 097 (2023)
[arXiv:2211.03668 [hep-ph]].

\bibitem{Grozin:2023dlk}
A.~Grozin,
JHEP \textbf{02}, 198 (2024)
[arXiv:2311.09894 [hep-ph]].

\bibitem{Grossman:2015cak}
Y.~Grossman, M.~K\"onig and M.~Neubert,
JHEP \textbf{04}, 101 (2015)
[arXiv:1501.06569 [hep-ph]].

\bibitem{Beneke:2023nmj}
M.~Beneke, G.~Finauri, K.~K.~Vos and Y.~Wei,
JHEP \textbf{09}, 066 (2023)
[arXiv:2305.06401 [hep-ph]].

\bibitem{Ishaq:2024pvm}
S.~Ishaq, S.~Zafar, A.~Rehman and I.~Ahmed,
PTEP \textbf{2024}, no.6, 063B05 (2024)
[arXiv:2404.01696 [hep-ph]].

\bibitem{Belle:2018jqd}
M.~Gelb \textit{et al.} [Belle],
Phys. Rev. D \textbf{98}, no.11, 112016 (2018)
[arXiv:1810.12976 [hep-ex]].

\bibitem{Bernlochner:2022qow}
F.~U.~Bernlochner, Z.~Ligeti, M.~Papucci, M.~T.~Prim, D.~J.~Robinson and C.~Xiong,
PoS \textbf{ICHEP2022}, 758 (2022)

\bibitem{Bernlochner:2022ywh}
F.~U.~Bernlochner, Z.~Ligeti, M.~Papucci, M.~T.~Prim, D.~J.~Robinson and C.~Xiong,
Phys. Rev. D \textbf{106}, no.9, 096015 (2022)
[arXiv:2206.11281 [hep-ph]].

\bibitem{Bernlochner:2023jkp}
F.~U.~Bernlochner, M.~Papucci and D.~J.~Robinson,
[arXiv:2312.07758 [hep-ph]].

\bibitem{HFLAV:Moriond 2024}
\url{https://hflav-eos.web.cern.ch/hflav-eos/semi/moriond24/r_dtaunu/RDRDst_hflav_moriond24_68.pdf}

\bibitem{Manzari:2024nxr}
C.~A.~Manzari and D.~J.~Robinson,
[arXiv:2402.12460 [hep-ph]].

\bibitem{Garg:2022bhj}
R.~Garg and A.~Upadhyay,
PTEP \textbf{2022}, no.9, 093B08 (2022)
[arXiv:2207.02498 [hep-ph]].

\bibitem{Vishwakarma:2022vzy}
K.~K.~Vishwakarma and A.~Upadhyay,
[arXiv:2208.02536 [hep-ph]].

\bibitem{Isgur:1989vq}
N.~Isgur and M.~B.~Wise,
Phys. Lett. B \textbf{232}, 113-117 (1989)
\end{thebibliography}
\end{document}